# Finite element and semi-analytical study of elastic wave propagation in strongly scattering polycrystals

Ming Huang[1,a], Peter Huthwaite[1], Stanislav I. Rokhlin[2] and Michael J. S. Lowe[1]

[1]Department of Mechanical Engineering, Imperial College London, Exhibition Road, London SW7 2AZ, United Kingdom

[2]Department of Materials Science and Engineering, Edison Joining Technology Center, The Ohio State University, 1248 Arthur E. Adams Drive, Columbus, OH 43221, United States

This work studies scattering-induced elastic wave attenuation and phase velocity variation in 3D untextured cubic polycrystals with statistically equiaxed grains using the theoretical second-order approximation (SOA) and Born approximation models and the grain-scale finite element (FE) model, pushing the boundary towards strongly scattering materials. The results for materials with Zener anisotropy indices $A > 1$ show a good agreement between the theoretical and FE models in the transition and stochastic regions. In the Rayleigh regime, the agreement is reasonable for common structural materials with $1 < A < 3.2$ but it deteriorates as $A$ increases. The wavefields and signals from FE modelling show the emergence of very strong scattering at low frequencies for strongly scattering materials that cannot be fully accounted for by the theoretical models. To account for such strong scattering at $A > 1$, a semi-analytical model is proposed by iterating the far-field Born approximation and optimising the iterative coefficient. The proposed model agrees remarkably well with the FE model across all studied materials with greatly differing microstructures; the model validity also extends to the quasi-static velocity limit. For polycrystals with $A < 1$, it is found that the agreement between the SOA and FE results is excellent for all studied materials and the correction of the model is not needed.

**Keywords:** elastic wave; polycrystal; strong scattering; attenuation and phase velocity; finite element; semi-analytical

## 1. Introduction

Elastic waves scatter as they travel through inhomogeneous media and thus exhibit scattering-induced attenuation and phase velocity dispersion. The problem of wave propagation and scattering in inhomogeneous media has received extensive study in the fields of, e.g., seismology [1,2] and non-destructive evaluation [3,4]. A subject of particular interest in both fields is to study wave propagation in polycrystalline media to facilitate the detection and characterisation of the inhomogeneities like faults, defects and grains within the media. The first complete theoretical treatment of the scattering-induced attenuation and velocity dispersion in polycrystals was conducted by Stanke and Kino [5] based on the multiple scattering theory developed by Karal and Keller [6,7]. This model was formulated for untextured polycrystals with statistically equiaxed grains of cubic symmetry. An equivalent model was later offered by Weaver [8] using the Dyson equation and the first-order smoothing approximation [9–11]. A variety of extensions of this later model has since been performed for various

---

ª Electronic mail: m.huang16@imperial.ac.uk

grain structures and crystal symmetries, see e.g. [12–15] or our earlier work [16–19] for an overview.

The theoretical models of the Stanke and Kino [5] and Weaver [8] type take the statistical information of the polycrystals as input and for a given wave modality they produce the scattering-induced attenuation coefficient and phase velocity as outputs. The models (before the Weaver model [8] invokes the Born approximation) are accurate when the second-order degree of inhomogeneity is small and hence we collectively call both models the second-order approximation (SOA) following Stanke and Kino [5]. The SOA results exhibit three specific frequency regions that are known as the Rayleigh, stochastic and geometric regimes with increased scattering intensities. All three regimes can also be predicted if the far-field approximation (FFA) [20] is invoked in the SOA model but the strongly scattering geometric regime vanishes if the single-scattering Born approximation is employed. The validity of these model approximations has recently been evaluated by 3D grain-scale finite element (FE) simulations which are capable of accurately describing the interaction of waves with grains [16–19,21–23]. These comparative studies demonstrated that the SOA, FFA and Born models agree very well with the FE results in the simulated Rayleigh and stochastic regimes. The SOA and FFA models are mostly indistinguishable from each other and have a better agreement with the FE results than the Born model. These studies showed that the theoretical models are valid for polycrystals with the spatial two-point correlation (TPC) of either scalar type for statistically equiaxed grains [16,17,21–23] or direction-dependent form for statistically elongated grains [18,19]. It was also shown that the elastic scattering factors [20], as combinations of elastic constants, are representative of the degree of inhomogeneity in the theoretical models for polycrystals of the highest cubic [16,18,19] and lowest triclinic [17–19] crystal symmetries; for cubic symmetry, the elastic scattering factors are related to the Zener anisotropy index $A$.

Despite the excellent performance of the SOA model, the comparative studies of the theoretical and FE models also revealed that, for cubic polycrystals with $A > 1$, the theoretical models start to deviate from the FE results at low frequencies as $A$ increases [12–15,18], while the agreement remains good in the transition and stochastic regimes. This is somewhat unexpected because it is reasonable to assume that the theoretical models perform less satisfactorily at high frequencies rather than at low frequencies since the degree of scattering increases with frequency. This finding has led us to hypothesise that strong scattering arises at low frequencies in strongly scattering polycrystals that is not fully considered by the theoretical models since they only account for a subset of scattering events. This work aims to further investigate this finding by studying a variety of equiaxed materials of cubic symmetry with greatly differing $A$; some studied materials have significantly larger $A$ than previously considered [5,16–19,21–23]. We shall see that the validity of the theoretical models is immediately challenged by the materials of very strong scattering, particularly, in the low-frequency range.

Consequently, it is desirable to develop a theoretical model valid for highly scattering materials. This subject of theoretical development has received extensive attention by first analysing a scatterer embedded in a homogeneous host medium followed by considering multiple scatterers, see e.g. [24,25] and the literature therein. However, this subject remains mostly unvisited for strongly scattering polycrystals. A few relevant studies include a recent 2D theory showing applicability to materials of high anisotropy [26] and a theory for strongly scattering materials [27] and the references therein. However, for longitudinal waves, only a small difference is found in Ref. [27] from the Stanke and Kino model [5]. To fill in this gap, this work proposes a semi-analytical model formulated iteratively from the far-field

Born approximation. The proposed model contains a second-order term on the elastic scattering factor and the iterative coefficient of this term is parametrically optimised for the model to best fit the FE results in the low-frequency range. We shall see that the proposed model works remarkably well in the Rayleigh, transition and stochastic regimes for various cubic polycrystals with largely differing anisotropies and greatly contrasting grain uniformities. The development of this semi-analytical model is to some extent empirical, but we expect that these promising simple closed-form solutions would stimulate future rigorous theoretical development.

Below, we first describe the 3D FE method in §2 and the theoretical models in §3, both for plane longitudinal wave propagation in equiaxed polycrystals of cubic symmetry. Then, we present a comparative study of the FE and theoretical results in §4 to evaluate the approximation of the theoretical models. Based on the evaluation of the results, we develop the semi-analytical model for strongly scattering materials and evaluate its applicability in §5. Conclusions are given in §6.

## 2. Finite element model

We use the 3D FE method to simulate the propagation of plane longitudinal waves in polycrystals with statistically equiaxed grains. We have reported the details of this method in our prior work [16,17,21,22,28] (see Refs. [18,19] for polycrystals with elongated grains) so only the essential steps are summarised below.

In the 3D FE method, we use a cuboid, composed of densely packed and fully bonded convex grains, to represent a polycrystal, as can be seen from Figure 1(a). The geometric models created for this work are detailed in Table 1. The selection of model dimensions ensures that there are more than 10 grains and 10 wavelengths in the z-direction of wave propagation. Each model is deployed in three separate forms, involving three microstructures with the same number of grains per unit volume but different grain uniformities, as illustrated in Figure 1(b). The three microstructures are generated using the Neper program [29] by the Laguerre, Poisson Voronoi, and centroidal Voronoi tessellations [29–31], as abbreviately called Laguerre, PVT and CVT hereafter. The PVT creates uniformly random seeds in the model space of a polycrystal, with each seed being enclosed by a convex grain within which all points are closer to the enclosed seed than to any other [16–19,21,22,28]. The equivalent spherical radii of the PVT grains are normally distributed as demonstrated in Refs. [21–23] and shown in Figure 1(c). In comparison to the PVT, the seeds of the Laguerre tessellation are weighted [30,31] to create the type of microstructure as commonly found in applications [32,33], in which case the equivalent grain radii follow the lognormal distribution as illustrated in Figure 1(c). The CVT iteratively changes the locations of the seeds to achieve a uniform distribution of grains and the resulting equivalent grain radii follow a much narrower normal distribution than that of the PVT [23], as can be seen from Figure 1(c). As discussed below and shown in Figure 1(d), the three microstructures have greatly differing TPC functions and thus induce distinctive scattering behaviours.

Each generated polycrystal model is then discretised in space and time. The spatial discretisation uses a structured mesh to divide the model space into identical 8-node linear 'brick' elements as illustrated in Figure 1(a). The element size $h$ of each model is listed in Table 1 and the size is chosen to be smaller than $\lambda/10$ and $D/10$ to satisfactorily suppress the numerical error of the simulation and to well represent the microstructure of the model [21,22,28], where $\lambda$ is the wavelength and $D$ is the cubic root of the average grain volume representing the average grain size. The time-stepping solution is based on the central difference scheme [28,34] using a time step of $\Delta t = 0.8h/V_{0L}$ that satisfies the Courant–Friedrichs–Lewy condition, where $V_{0L}$ is the longitudinal Voigt velocity.

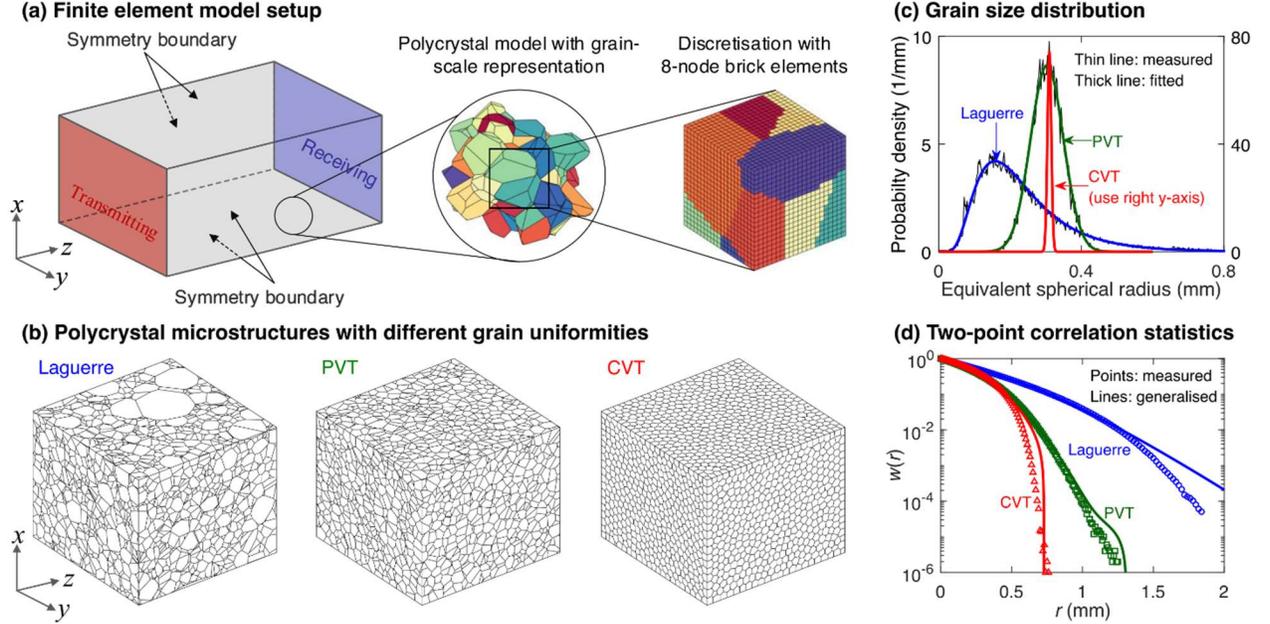

Figure 1. (a) Finite element model setup for the simulation of plane longitudinal wave propagation in polycrystals with statistically equiaxed grains. (b) Three polycrystal microstructures of different grain uniformities; each illustrated microstructure has total dimensions of 12×12×10 mm and includes 11520 grains. (c) Grain size distributions for the three microstructures, represented by the probability density of equivalent spherical grain radii; the fitted distribution for Laguerre is lognormal while those for PVT and CVT are normal; the respective mean values are 0.24, 0.30 and 0.31 mm and standard deviations are 0.13, 0.05 and 0.005 mm. (d) Two-point correlation statistics for the three polycrystal microstructures.

Each polycrystal model is subsequently assigned with material properties. We consider single-phase polycrystals in this work, so the individual grains of each model have the same mass density and elastic constants. The grains within each model are defined with uniformly randomly oriented crystallographic axes, making the model macroscopically homogeneous and isotropic (untextured) [21]. The polycrystalline materials used in this work are recorded in Table 2. The materials have the same cubic crystal symmetry but different anisotropy indices $A = 2c_{44}/(c_{11} - c_{12})$, with eight materials having $A > 1$ (four naturally occurring and four fictitious [16]) and four having $A < 1$ [35].

We simulate the propagation of plane longitudinal waves in the $z$-direction. To initiate this wave modality, symmetry boundary conditions (SBCs) are defined for the four lateral outer surfaces of each model, i.e. $x = 0$, $x = d_x$, $y = 0$ and $y = d_y$ surfaces in Figure 1(a). In applying the SBCs, we have the displacement components of $u_x(x, y, z) = -u_x(-x, y, z)$, $u_y(x, y, z) = u_y(-x, y, z)$ and $u_z(x, y, z) = u_z(-x, y, z)$ for the $x = 0$ surface for example. This essentially means that the $x = 0$ surface acts as a mirror that reflects the model to form a virtual symmetric model on the other side of the surface and is physically equivalent to the condition of $u_x(x = 0, y, z) = 0$ on the $x = 0$ surface. When combined with the SBCs on the $x = d_x$ surface, the model would be repeatedly reflected on the $x = 0$ and $x = d_x$ surfaces and their subsequent mirrored surfaces, leading to a virtually infinitely wide model in the $x$-direction. Together with the SBCs on the $y = 0$ and $y = d_y$ surfaces, an infinitely wide model would be formed across the $x$-$y$ plane to sustain plane longitudinal waves. We note that a practically equivalent boundary condition, namely the periodic boundary condition [22], can be used in this case, but we refrain from using it here because it substantially increases the complexity of the FE solution [22]. To excite the desired $z$-direction plane longitudinal wave, a $z$-direction force in the form of a three-cycle Hann-windowed toneburst is uniformly applied to every node on the $z = 0$ surface.

Table 1. Polycrystal models. Dimensions $d_x \times d_y \times d_z$ (mm), number of grains $N$, average grain diameter $D$ (mm, cubic root of average grain volume), mesh size $h$ (mm), degree of freedom d.o.f., frequency range $f$ (MHz) for aluminium modelling and the respective number of wavelengths $n_\lambda = d_z/(V_{0L}/f)$ per model dimension in the z-direction of wave propagation (these numbers vary slightly for the modelling of other materials).

| Model | $d_x \times d_y \times d_z$ | $N$ | $D$ | $h$ | d.o.f. | $f$ | $n_\lambda$ |
|---|---|---|---|---|---|---|---|
| N115200 | 12×12×100 | 115200 | 0.5 | 0.050 | 349×10⁶ | 1.0 – 6.5 | 16 - 103 |
| N11520 | 12×12×10 | 11520 | 0.5 | 0.025 | 278×10⁶ | 6.5 - 13.5 | 10 - 21 |
| N16000 | 20×20×5 | 16000 | 0.5 | 0.020 | 755×10⁶ | 13.5 - 25.0 | 11 - 20 |

Table 2. Polycrystalline materials with cubic crystal symmetry. Elastic constants $c_{ij}$ (GPa), density $\rho$ (kg/m³), Voigt velocities $V_{0L/T}$ (m/s), Zener anisotropy index $A$, elastic scattering factors $Q_{L\rightarrow L/T}$.

| | $c_{11}$ | $c_{12}$ | $c_{44}$ | $\rho$ | $V_{0L}$ | $V_{0T}$ | $A$ | $Q_{L\rightarrow T}$ | $Q_{L\rightarrow L}$ |
|---|---|---|---|---|---|---|---|---|---|
| Aluminium | 103.4 | 57.10 | 28.60 | 2700 | 6318 | 3128 | 1.24 | 3.34×10⁻⁴ | 7.80×10⁻⁵ |
| A=1.5 | 262.1 | 136.5 | 95.30 | 8000 | 6001 | 3207 | 1.52 | 1.43×10⁻³ | 3.88×10⁻⁴ |
| A=1.8 | 251.7 | 141.7 | 100.5 | 8000 | 6001 | 3207 | 1.83 | 2.79×10⁻³ | 7.60×10⁻⁴ |
| A=2.4 | 237.1 | 149.0 | 107.8 | 8000 | 6001 | 3207 | 2.45 | 5.48×10⁻³ | 1.49×10⁻³ |
| Copper | 169.6 | 122.4 | 74.00 | 8935 | 4847 | 2455 | 3.14 | 7.19×10⁻³ | 1.76×10⁻³ |
| Inconel | 234.6 | 145.4 | 126.2 | 8260 | 6025 | 3366 | 2.83 | 7.59×10⁻³ | 2.26×10⁻³ |
| A=5.0 | 210.6 | 162.1 | 121.0 | 8000 | 6000 | 3207 | 5.00 | 1.27×10⁻² | 3.44×10⁻³ |
| Lithium | 13.40 | 11.30 | 9.600 | 534.0 | 6157 | 3402 | 9.14 | 1.87×10⁻² | 5.44×10⁻³ |
| RbF | 55.30 | 14.00 | 9.300 | 3557 | 3605 | 1973 | 0.45 | 6.44×10⁻³ | 1.84×10⁻³ |
| RbCl | 36.30 | 6.200 | 4.700 | 2760 | 3186 | 1790 | 0.31 | 1.38×10⁻² | 4.16×10⁻³ |
| RbBr | 31.70 | 4.200 | 3.880 | 3349 | 2666 | 1529 | 0.28 | 1.67×10⁻² | 5.24×10⁻³ |
| RbI | 25.80 | 3.700 | 2.800 | 3550 | 2326 | 1311 | 0.25 | 1.86×10⁻² | 5.63×10⁻³ |

Each polycrystal model is solved in the time domain using the GPU-accelerated Pogo program [34]. $z$-direction displacements are monitored during the time-stepping solution, and example results are provided in Figure 2 for a single realisation of the model N115200 with the PVT microstructure, simulated at a centre frequency of 1 MHz ($2k_{0L}a \approx 1$). Figure 2(a) shows the displacement wavefields on an arbitrary cross-section at an arbitrary time. Figure 2(b) displays the signals at individual nodes as thin grey lines and shows the respective coherent signals as thick lines that are averaged over all nodes on the monitoring boundaries. Figure 2(c) presents the normalised wavefront fluctuations on the receiving boundary at a normalised frequency of $2k_{0L}a = 1$. To find the value of such a fluctuation, the time-domain displacement field at $z = d_z$, namely $u_z(x, y; t)$, is Fourier transformed to the frequency domain, $u_z(x, y; f)$. The amplitude of the resulting field is normalised by the coherent amplitude to get the fluctuation as $u_f(x, y; f) = |u_z(x, y; f)|/\langle |u_z(x, y; f)| \rangle_{x,y} - 1$, where $\langle \cdot \rangle_{x,y}$ represents the coherent average over all $x$ and $y$ nodes. The fluctuations can be alternatively given in the time domain but they are not provided here because they essentially offer the same information as the frequency-domain ones, see Ref. [28] for details. Since the fluctuation is normalised by the coherent amplitude, its root-mean-square (RMS) is essentially the normalised standard deviation and thus quantifies the uncertainty of the fluctuation; the RMS values are annotated in Figure 2(c).

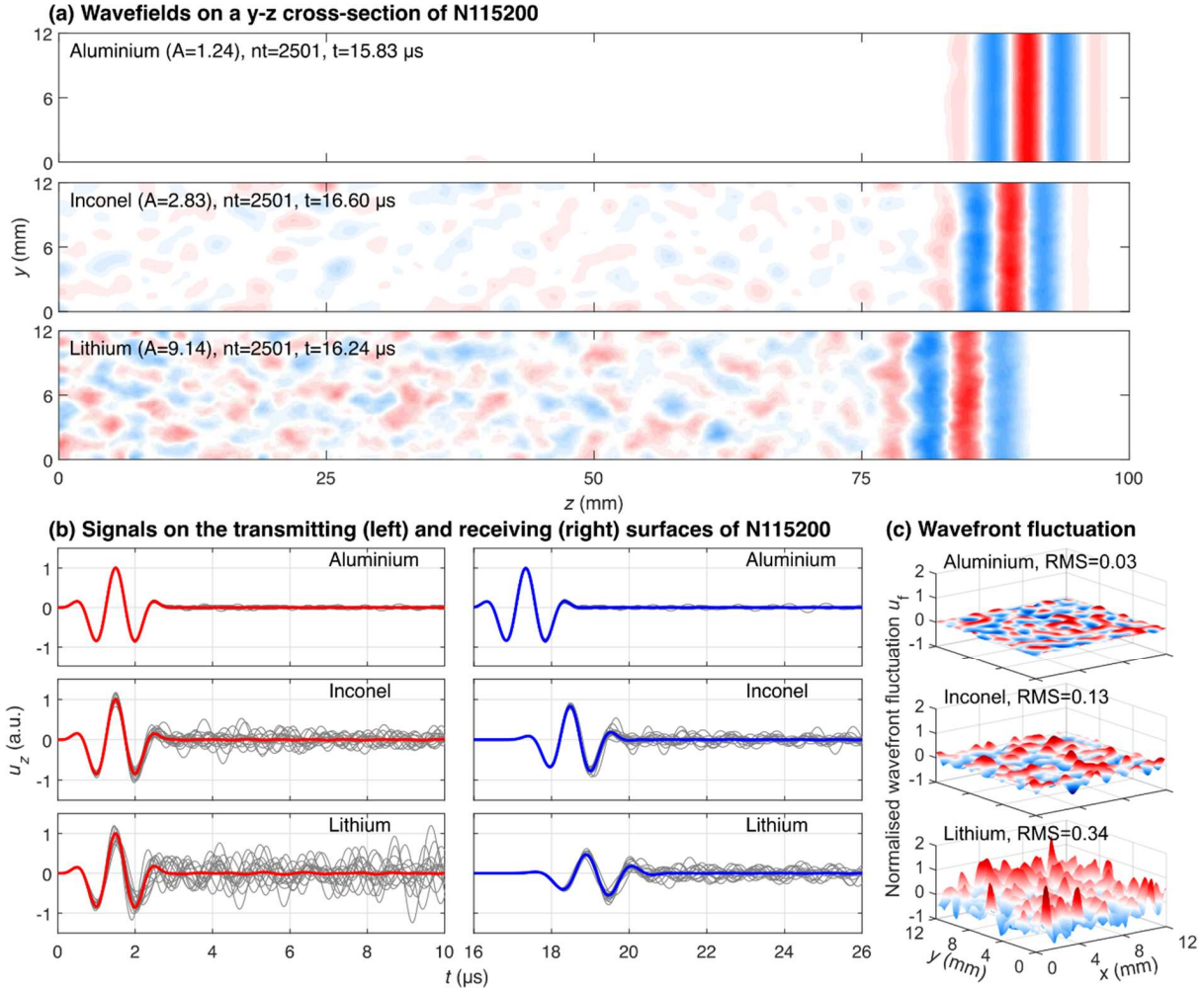

Figure 2. (a) Cross-sectional wavefields, (b) transmitted and received signals and (c) normalised wavefront fluctuations for plane longitudinal waves in aluminium, Inconel and lithium, simulated at a centre frequency of 1 MHz ($2k_{0L}a \approx 1$) using a single realisation of the model N115200 with the PVT microstructure. The wavefields in (a) are given for an arbitrary cross-section at an arbitrary time. The thin grey lines in (b) represent the signals monitored at individual nodes on the transmitting and receiving surfaces, while the thick lines are the coherent signals averaged over all nodes. The normalised wavefront fluctuations in (c) are given for the receiving surface at $2k_{0L}a = 1$ and the root-mean-square (RMS) of fluctuation is annotated; colourmap is illustrative only.

All figure panels show stronger scattering as $A$ increases from aluminium through Inconel to lithium, leading to weaker coherent signals on the receiving boundary. However, the recorded coherent signals are strong and clear even for lithium. The coherent waves are only affected by the multiple scattering events arriving simultaneously with the main coherent beam. In addition, the incoherent waves are diminished after averaging over a significant number (~60,000 in the least case) of monitoring nodes in each case;

this was discussed in-depth in Sec. IV.3 of Ref. [19]. Therefore, the coherent signals have a high degree of signal-to-noise ratio (SNR) and the SNR is further improved in this work by averaging over as many as 15 repeated realisations of the model domain as will be discussed below. Such coherent signals are well suited to calculating scattering induced attenuation and velocity change. This is achieved by applying appropriate time windowing to the emitted and received coherent signals and then comparing the resulting main wave pulses in the frequency domain; see Refs. [21,22,28] for details. The SNR would decline as frequency increases towards the transition to the geometric regime, but all FE results reported in this work have achieved a good SNR by limiting them to a reasonable frequency range [19].

We note that we use all three models in Table 1 for the simulation of each material in Table 2. The individual models are simulated with different centre frequencies for the applied toneburst force, allowing each material to be simulated in a wide frequency range as reported later in this work; example frequency ranges covered by individual models are provided in Table 1 for aluminium. The combination of different polycrystal models with various centre frequencies delivers a very high degree of numerical accuracy across the whole frequency range [28]. Besides, the simulation results presented in this work have achieved a very good statistical convergence by taking the average of 15 realisations for each case (each model-frequency combination). The different realisations of each case use the same polycrystal model but the crystallographic orientations of the grains are randomly reshuffled for each realisation [22]. Alternatively, these multiple realisations can also be based on polycrystal models of different grain arrangements [22]. It was shown [22] that this approach is equivalent to, but more computationally intensive than, that used here. The applicability of our approach is also supported by the FE results presented in §4 and §5 that show consistency across different models at their overlapping frequencies for all studied materials of different microstructures.

The TPC statistics $w(r)$, describing the probability of two points separated by a distance of $r$ falling into the same grain, are numerically measured from the generated polycrystal models and the resulting data points are shown in Figure 1(d) for the three polycrystal microstructures. Here we treat the TPC as a scalar function because it is direction independent. Taking the model N11520 with the PVT microstructure as an example, the TPC curves measured in 30 randomly chosen directions have a mean correlation length (i.e., integral of TPC curve) of 0.23 mm and a very small standard deviation of $7.19 \times 10^{-4}$ mm. This supports the direction independence of the TPC and also substantiates the statistically equiaxed nature of the grains. Since all three polycrystal models in Table 1 have a large number of grains, they possess nearly the same statistical characteristics for a given microstructure. Taking the PVT microstructure as an example, the TPCs of the three models are indistinguishable [28] and their average correlation length is 0.23 mm, with a standard deviation of $5.84 \times 10^{-5}$ mm. For this reason, we have selected the model N11520 to determine the TPC for each microstructure. We note that the SBCs effect, which doubles the sizes of the symmetry boundary grains and makes these grains larger on average than the grains located within the body of the models [28], is not considered when measuring the TPC; this will be further discussed in §4(a). To incorporate the measured statistics into the theoretical models, they are fitted into generalised TPC functions, $w(r) = \sum_i A_i e^{-r/a_i}$, which are displayed in Figure 1(d) as solid lines; the $A_i$ and $a_i$ coefficients are provided in Supplementary Table 1 for the three microstructures. Detailed TPC measurement and fitting procedures were reported in Refs. [16,17,28]. As can be found in Figure 1(d), the agreement of the generalised TPC curves with the points is less satisfactory at the tail than at the origin. Improving the agreement at the tail would increase the accuracy of the volumetric characteristic of the polycrystal, as described by the effective grain volume $V_{\text{eff}}^{\text{g}} = \int w(r) \mathrm{d}r^3 = 8\pi \sum_i (A_i a_i^3)$ [5,8,17] in Supplementary Table 2. However, the accuracy improvement is very limited and is thus not pursued here because the TPC at the tail is very small in probability. In Supplementary Table 2, we also provide the other two characteristic parameters of the TPC as will be used below, including the mean line intercept $a = -1/w'(r=0) = 1/\sum_i(A_i/a_i)$ [5,17,36] and correlation length $a_{\text{CL}} = \int_0^\infty w(r)\mathrm{d}r = \sum_i A_i a_i$ [5,37,38]. Note that the three microstructures have approximately the same mean line intercept (i.e., the same slope at the origin).

## 3. Theoretical models

As with the above FE method, the same wave propagation problem of elastic wave propagation in polycrystals with statistically equiaxed grains is addressed here from a theoretical perspective. The theoretical models considered here use the statistical information of the FE models as input, enabling a direct comparison of both methods. These models are briefly introduced below; readers are referred to Ref. [17] for details.

***Second-order approximation (SOA):*** The equiaxed polycrystals considered in this work are macroscopically homogeneous and isotropic, the spatially-varying elastic tensor of a polycrystal can thus be expressed as $c_{ijkl}(\mathbf{x}) = c_{ijkl}^0 + \delta c_{ijkl}(\mathbf{x})$, with the Voigt average $c_{ijkl}^0 = \langle c_{ijkl}(\mathbf{x}) \rangle$ representing the homogeneous reference medium and $\delta c_{ijkl}(\mathbf{x})$ denoting the elastic fluctuation. An incident wave scatters on the elastic fluctuation and the wave number is thus perturbed as the wave propagates. The perturbed wave number $k$ satisfies the dispersion equation [8,16,17,20]

$$\omega^2 - k^2 V_{0M}^2 - m_M(\mathbf{k}; \omega) = 0, \qquad (1)$$

where $\mathbf{k} = k\mathbf{p}$ is the wave vector; the unit vector $\mathbf{p}$ represents the wave propagation direction, which is arbitrary due to the macroscopic isotropy of the polycrystal. $\omega = 2\pi f$ is the angular frequency and $f$ is the frequency. $V_{0M}$ represents the Voigt phase velocity of the wave $M$ in the homogeneous reference medium. $m_M = \sum_{N=\text{L,T}} m_{M \to N}$ is the spatial Fourier transform of the mass operator describing the random scattering events occurring in the polycrystal. The component $m_{M \to N}$, denoting the scattering of the wave $M$ into $N$, is given below [16,17,20] by using the first-order smoothing approximation [8,9] (equivalent to the Bourret approximation [9–11])

$$m_{M \to N}(\mathbf{k}; \omega) = \frac{2\pi k^2 k_{0N}^3}{\eta \rho^2 V_{0N}^2} \left\{ \mathrm{P.V.} \int_0^\infty \left[ \frac{\xi^4}{1-\xi^2} \int_0^\pi f_{M \to N}(k, \omega, \xi, \theta) \sin\theta \mathrm{d}\theta \right] \mathrm{d}\xi \right. \\ \left. -i\frac{\pi}{2} \int_0^\pi f_{M \to N}(k, \omega, \xi = 1, \theta)\sin\theta \mathrm{d}\theta \right\}, \qquad (2)$$

where the mass density $\rho$ is constant for a single-phase polycrystal considered in this work. $k_{0N}$ denotes the wave number of the wave $N$ in the reference medium. P.V. represents the Cauchy principal value and $\xi$ is a dimensionless variable. The coefficient $\eta$ is 1 and 2 for longitudinal ($M = \text{L}$) and transverse ($M = \text{T}$) propagating waves, respectively. The factor $f_{M \to N}$ in Eq. (2) describes the TPC of the elastic fluctuation and is given by [16,17,20]

$$f_{M \to N}(k, \omega, \xi, \theta) = (A_{MN} + B_{MN}\cos^2\theta + C_{MN}\cos^4\theta) \sum_i \frac{A_i a_i^3}{\pi^2 \left[1 + a_i^2 \left(k^2 + \xi^2 k_{0N}^2 - 2\xi k k_{0N}\cos\theta\right)\right]^2}, \qquad (3)$$

where the terms in the summation symbol correspond to the spectral representation of the spatial TPC function, $w(r) = \sum_i A_i e^{-r/a_i}$, and the rest terms in the parentheses represent the elastic part of the TPC. For longitudinal waves in cubic polycrystals, the $A_{MN}, \ldots$ coefficients in Eq. (3) ($M =$ L, $N \in \{$L, T$\}$) are given by $A_{LL} = 3c^2/175$, $A_{LT} = c^2/35$, $B_{LL} = B_{LT} = 2c^2/175$, and $C_{LL} = -C_{LT} = c^2/525$, where $c = c_{11} - c_{12} - 2c_{44}$ is the invariant anisotropy coefficient [16] ($c = 0; A = 1$ for isotropy); those coefficients for arbitrary crystal symmetries can be found in our prior work [17].

One obtains the perturbed wave number $k$ for the propagating wave $M$ by numerically solving the dispersion equation, Eq. (1). Consequently, the attenuation coefficient and phase velocity are calculated from the imaginary and real parts of the perturbed wave number by $\alpha_M = \text{Im} k$ and $V_M = \omega/\text{Re} k$.

***Far-field approximation (FFA):*** The FFA model does not involve the complex calculation of the Cauchy integral as in the SOA model and it has an explicit expression for the mass operator as given below for polycrystals with statistically equiaxed grains [17,20]

$$m_{M \to N} = \sum_i \frac{-4A_i k^2 a_i^2 V_{0M}^2 k_{0N}^2 Q_{M \to N}}{k^2 a_i^2 - (i + k_0 N a_i)^2}. \tag{4}$$

Similarly to the SOA model, substituting Eq. (4) into the dispersion equation, Eq. (1), and numerically solving the equation results in the solution to the perturbed wave number $k$. We note that the resulting velocity, $V_M = \omega/\text{Re} k$, needs to be corrected by adding a constant velocity of $V_M^R - V_{0M}$ [17,20], with $V_M^R$ being the Rayleigh velocity limit given below. The solution of the FFA model is mostly indistinguishable from that of the SOA model across the whole frequency range but the former is more computationally efficient [17]. An important advantage of the FFA model for our purpose is that it explicitly relates the attenuation and velocity dispersion to the elastic scattering factors $Q_{M \to N}$ [20]. Two factors, $Q_{M \to M} = (A_{MM} + B_{MM} + C_{MM})/(4\eta \rho^2 V_{4M}^4)$ and $Q_{M \to N} = (A_{MN} + B_{MN}/3 + C_{MN}/5)/(4\eta \rho^2 V_{0M}^2 V_{0N}^2)$ ($N \neq M$) [20], exist for macroscopically isotropic polycrystals with grains of arbitrary symmetry, and they have simple expressions of $Q_{L \to L} = 4c^2/(525\langle c_{11}\rangle^2)$ and $Q_{L \to T} = c^2/(105\langle c_{11}\rangle\langle c_{44}\rangle)$ for longitudinal waves in a cubic polycrystal [20].

***Born approximation:*** Closed-form solutions can be found for the SOA and FFA models by invoking the Born approximation. Since the resulting numerical solutions are nearly the same [17], here we only provide the Born approximation of the FFA model. To obtain the solution, we substitute $(\omega/V_{0M})^2 - k^2$ by $2k_{0M}(k_{0M} - k)$ in Eq. (1) and replace $\mathbf{k}$ with $\mathbf{p}k_{0M}$ in Eq. (4). This leads to the following solution for a longitudinal propagating wave [17,20]

$$k_L = k_{0L} + 2k_{0L}^3 Q_{L \to L} \sum_i \frac{A_i a_i^2}{k_{0L}^2 a_i^2 - (i + k_{0L} a_i)^2} + 2k_{0L} k_{0T}^2 Q_{L \to T} \sum_i \frac{A_i a_i^2}{k_{0L}^2 a_i^2 - (i + k_{0T} a_i)^2} + 2(Q_{LL}^* + Q_{L \to T})k_{0L}, \tag{5}$$

where $2(Q_{LL}^* + Q_{L \to T})k_{0L}$ is added to consider the above-mentioned velocity correction. The imaginary and real parts of the solution $k_L$ are further obtained as

$$\alpha_L = \text{Im} k_L = \sum_i A_i \frac{4Q_{L \to L} k_{0L}(k_{0L} a_i)^3}{1 + 4(k_{0L} a_i)^2} + \sum_i A_i \frac{4Q_{L \to T} k_{0L}(k_{0T} a_i)^3}{\left[1 + (k_{0T} a_i)^2(\eta_{LT}^2 - 1)\right]^2 + 4(k_{0T} a_i)^2}, \tag{6}$$

$$\text{Re} k_L = k_{0L} + \sum_i A_i \frac{2Q_{L \to L} k_{0L}(k_{0L} a_i)^2}{1 + 4(k_{0L} a_i)^2} + 2Q_{LL}^* k_{0L}$$
$$+ \sum_i A_i \frac{2Q_{L \to T} k_{0L}(k_{0T} a_i)^2\left[1 + (k_{0T} a_i)^2(\eta_{LT}^2 - 1)\right]}{\left[1 + (k_{0T} a_i)^2(\eta_{LT}^2 - 1)\right]^2 + 4(k_{0T} a_i)^2} + 2Q_{L \to T} k_{0L}, \tag{7}$$

where $\eta_{LT} = V_{0T}/V_{0L}$.

***Rayleigh asymptotes:*** At the low-frequency Rayleigh limit, the attenuation and phase velocity asymptotes are given by [17]

$$\alpha_M^R = \frac{1}{2\pi} k_{0M}^4 V_{\text{eff}}^g \left( Q_{MM}^* + \frac{V_{0M}^3}{V_{0N}^3} Q_{M\to N} \right), \quad V_M^R = \frac{V_{0M}}{1 + 2Q_{MM}^* + 2Q_{M\to N}} \tag{8}$$

where $N \neq M$. $V_{\text{eff}}^g$ is the effective grain volume defined by the volumetric integral of the TPC function [5,8,17], Supplementary Table 2. $Q_{MM}^* = (A_{MM} + B_{MM}/3 + C_{MM}/5)/(4\eta\rho^2 V_{0M}^4)$ is an elastic factor for simplifying the equation. For longitudinal waves, since $Q_{LL}^*$ is generally negligible and $V_{0L}^3/V_{0T}^3$ just differs slightly among most structural materials, we can obtain from Eq. (8) that $\alpha_L^R \propto Q_{L\to T}$. Appending the fact that $Q_{LL}^*$ and $Q_{L\to T}$ are far smaller than unity, it follows from Eq. (8) that the phase velocity variation is $V_L^R/V_{0L} - 1 \approx -2Q_{LL}^* - 2Q_{L\to T} \approx -2Q_{L\to T}$. Therefore, it is sufficient to use $Q_{L\to T}$ to characterise the level of scattering for the materials in the Rayleigh regime.

***Stochastic asymptotes:*** At the high-frequency stochastic limit, the attenuation and phase velocity asymptotes are given by [17]

$$\alpha_M^S = k_{0M}^2 a_{\text{CL}} Q_{M\to N}, \quad V_M^S = \frac{V_{0M}}{1 + 5Q_{M\to N}/2 + 2Q_{MN}^*/(1 - V_{0M}^2/V_{0N}^2)} \tag{9}$$

in which $N \neq M$. $a_{\text{CL}}$ is the correlation length defined by the integral of the TPC function [37], Supplementary Table 2. $Q_{MN}^* = (A_{MN} + B_{MN} + C_{MN})/(4\eta\rho^2 V_{0M}^2 V_{0N}^2)$ is an elastic factor introduced for simplifying the equation. For longitudinal waves, it follows from Eq. (9) that $\alpha_L^S \propto Q_{L\to L}$ and $V_L^S/V_{0L} - 1 \approx -5Q_{L\to L}/2 + 2Q_{LT}^*(V_{0L}^2/V_{0T}^2 - 1) \propto Q_{L\to L}$. Thus, $Q_{L\to L}$ is the major elastic factor determining the level of scattering at the stochastic limit.

## 4. Comparison of finite element and theoretical models

Now we present and discuss the numerical FEM results and theoretical predictions. The first three subsections focus on the eight cubic materials with anisotropy indices $A > 1$, while the last subsection on the four materials with $A < 1$. Both cases consider the PVT microstructure only.

### (a) Dependence of attenuation and velocity on frequency

Figure 3 shows the normalised attenuation $2\alpha_L a$ and the phase velocity variation $V_L/V_{0L} - 1$ versus the normalised frequency $2k_{0L}a$ for the eight $A > 1$ materials (Table 2) with the PVT microstructure. The normalisation factor $a$ is the mean line intercept of the grains, Supplementary Table 2. Points are FEM results and curves are theoretical predictions.

The FEM points of each material are obtained using all three polycrystal models in Table 1, modelled with multiple centre frequencies, to cover the shown frequency range. A combination of 15 realisations is used for each modelling case to achieve statistically meaningful results. The average of the multiple realisations is shown as the points in the figure while the corresponding standard deviation (i.e., error bar) is not provided because it is smaller than the size of the FEM point markers. The relative differences between the theoretical curves and the FEM points are provided in Figure 4. A small discontinuity (sudden jump) can be observed as we look along sequential FEM points for each material, which is more evident from the phase velocity points for A=5.0 and lithium at around $2k_{0L}a = 1$ (those jumps are especially visible for the relative differences in Figure 4). Such a discontinuity occurs because two different FE material models are used for calculating its left- and right-side FEM points; e.g. N115200 and N11520 for the left and right sides of $2k_{0L}a \approx 1$. The models on the two sides have different mesh sizes and thus different numbers of elements per wavelength at their overlapping frequency of calculation. This causes different numerical errors [28] that exhibit a discontinuity of the FEM results. The discontinuity is observable in the phase velocity results because the FE scheme used in this work is more prone to numerical phase errors, namely numerical dispersion [28]. Also, this discontinuity is more evident for highly scattering materials because numerical errors depend on material anisotropy [28]. The discontinuities essentially define the bound of the numerical error, which is mostly one order of magnitude smaller than the difference between the theoretical and numerical FE results, as can be more evidently seen from Figure 4. For the extreme case, i.e. the phase velocity of lithium, the numerical error is about 0.4%, whereas the SOA-FEM difference is 2.4%. In essence, the FE results have achieved a very high degree of numerical accuracy and statistical convergence (also see our prior studies [21,22,33] and especially [28] for a more detailed assessment of these two aspects), and therefore they are used as reliable references below to evaluate the approximations of the theoretical models.

The theoretical SOA and Born curves are produced by incorporating the generalised TPC function of the FE material models, Supplementary Table 1. The FFA model results are indistinguishable from the SOA curves [17] and are thus not provided. We note that the SBCs effect is not considered in the measured and generalised TPC functions and is accordingly not accounted for in the theoretical SOA and Born curves. Since the SBCs enlarge the symmetry boundary grains and slightly increase the scattering intensity in the FE simulations (such increase occurs to both the L-L and L-T scattering components but the L-T component may increase more at low frequencies where the L-T component is dominant) [28], the theoretical predictions would underestimate the level of attenuation and phase velocity variation in comparison to the FE results. Nonetheless, our prior work [28] has shown that this underestimation is very small and the same estimation for all materials addressed in this work reveals that this underestimation does not depend on material anisotropy. Also, it can be seen from Figure 2(a) that there is no visible distortion of the waves on the symmetry boundaries for all materials, thus supporting the smallness and anisotropy independence of the SBCs effect. For this reason, we ignore the SBCs effect in this work.

For attenuation, the theoretical results approach the Rayleigh and stochastic asymptotes in the low- and high-frequency regions for all materials, and therefore they have fourth- and second-power dependences on frequency in the two regions, as follows from Eqs. (8) and (9); the FEM results display the same frequency dependencies. The 'hump' in the transition frequency regime is a unique feature for longitudinal waves, which is attributed to the transition of the dominant $L \rightarrow T$ scattering in the Rayleigh regime to the dominant $L \rightarrow L$ in the stochastic regime [5,17,23]. Also, for a more strongly scattering material [5,17,19] the attenuation results show a less pronounced stochastic region and an earlier transition to the geometric regime (seemingly flat part appearing at the far right of the SOA curve for lithium). For phase velocity, the FEM and theoretical results tend to be non-dispersive in the Rayleigh regime while having different static limits for high anisotropies. Between the Rayleigh and stochastic asymptotes the phase velocity is dispersive due to transition of the dominant scattering mechanisms. The velocity results also show a shorter stochastic regime for a more anisotropic material, and this regime can hardly be observed from the FEM points because (1) for weakly scattering materials (aluminium, A=1.5, A=1.8 and A=2.4), the velocity decreases slowly with frequency due to the numerical dispersion in the FE simulation [28], while (2) for strongly scattering materials, the velocity increases with frequency potentially as a result of the numerical dispersion and the complete disappearance of the stochastic regime due to the strongly scattering nature of the materials.

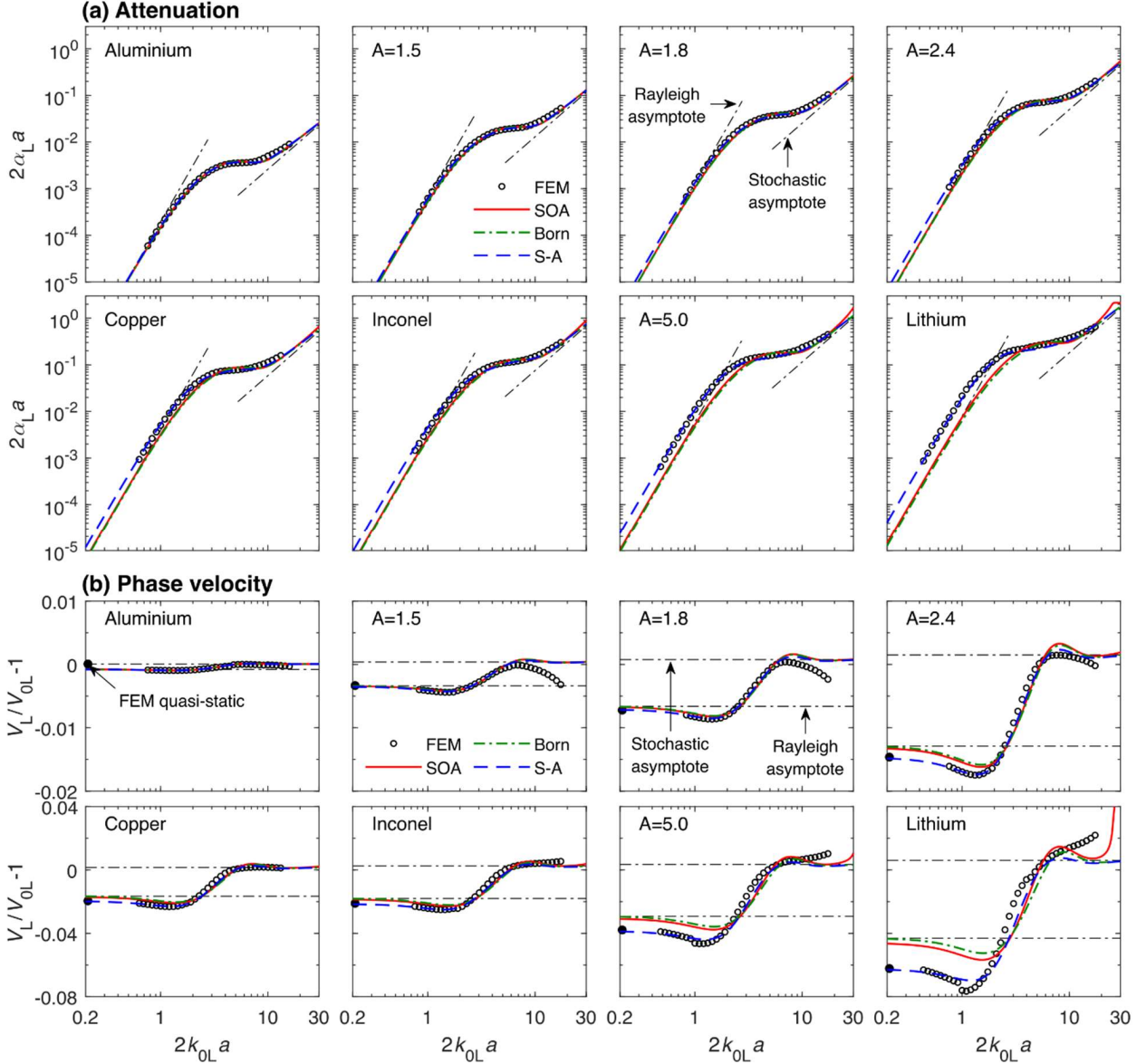

Figure 3. (a) Normalised attenuation and (b) phase velocity variation versus normalised frequency for plane longitudinal waves in eight cubic polycrystals with $A > 1$ (the PVT microstructure). The figure shows a comparison of numerical FEM results (points) with theoretical SOA and Born, and semi-analytical (S-A) (dashed lines) predictions. All FEM points are obtained by averaging the results of 15 FE simulations; the corresponding error bars are not shown since they are smaller than the size of the point markers. The theoretical SOA and Born, and semi-analytical curves have incorporated the generalised TPC function (Supplementary Table 1) of the FE models. The elastic scattering factors $Q_{L \to T}$ and $Q_{L \to L}$ of the materials, Table 2, follow the sequence: aluminium < A=1.5 < A=1.8 < A=2.4 < copper < Inconel < A=5.0 < lithium. The leftmost solid points in (b) are quasi-static FEM results. The y-axis range of the top four panels in (b) is one-fourth of that of the bottom four.

Considering both the low- and high-frequency regimes, the attenuation and velocity variation ranges in Figure 3 increase with the material anisotropy. This is due to the increase of scattering with material anisotropy, as can be observed from the wavefields, signals and wavefront fluctuations of different materials in Figure 2. Quantitative evidence can also be found from the fluctuation RMS values in Figure 2(c), which show a tenfold increase from aluminium to lithium. This increase of the fluctuation RMS can be further characterised by a quadratic relationship to the scattering factor $Q_{L \to T}$ by RMS = $206.96Q_{L \to T}^2 + 12.18Q_{L \to T} + 0.04$, with a goodness-of-fit of $R^2 = 0.987$. This quadratic fit is generated using the three RMS values in Figure 2(c) as well as those not shown, which are 0.05, 0.09, 0.11, 0.15 and 0.22 for the materials A=1.5, A=1.8, A=2.4, copper and A=5.0.

Following the increase of scattering with material anisotropy, the theoretical SOA and Born curves in Figure 3 show a deteriorated agreement with the FEM points as material anisotropy increases. As aforementioned, this is because the theoretical models only account for a subset of scattering events whereas the FEM points accurately incorporate all possible events. A further quantitative evaluation is provided in Figure 4, showing the relative differences between the theoretical and FEM results. In the figure, the relative difference in attenuation between the SOA and FEM results, for instance, is calculated by $\delta = \alpha_L^{SOA}/\alpha_L^{FEM} - 1$, with the SOA and FEM results $\alpha_L^{SOA}$ and $\alpha_L^{FEM}$ taken from Figure 3.

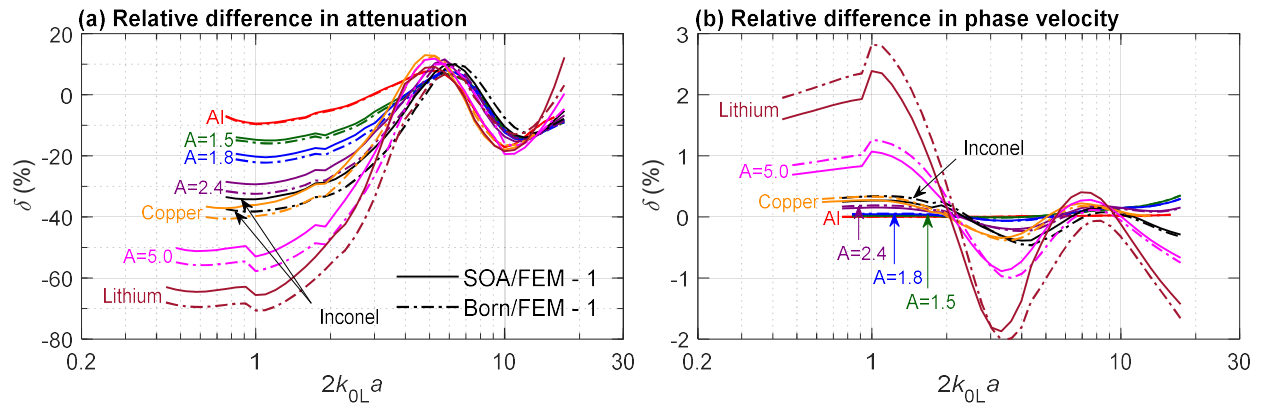

Figure 4. Relative difference in (a) attenuation and (b) phase velocity between theoretical (SOA and Born) and numerical (FEM) results, plotted against normalised frequency for longitudinal waves in equiaxed cubic polycrystals of the PVT microstructure. All theoretical and FEM results are taken from Figure 3. Their relative differences, with the FEM results as the reference, are shown in percentage.

A distinctive observation from Figure 4 is that the relative differences in attenuation and phase velocity between the theoretical and FE models are relatively larger at low frequencies. Taking the frequency at $2k_{0L}a = 1$ as an example, the relative difference in attenuation between the SOA and FEM results increases from -10% for aluminium to -66% for lithium, while the relative difference in phase velocity increases from $4 \times 10^{-4}\%$ for aluminium to 2.4% for lithium. Such a large SOA-FEM difference is somewhat unexpected since it was previously believed that scattering is small in the low-frequency range and can thus be appropriately accounted for by the SOA model. However, it can be observed from Figure 2 that very strong scattering arises at low frequencies, especially for strongly scattering lithium. Also, a strong reverberating signal is visible in Figure 2(b) long after the main pulse, unquestionably indicating the existence of strong multiple scattering. One of the limitations of the SOA model is that it considers multiple scattering only partially, and the Born approximation accounts for only single scattering events and thus deviates even more greatly from the FEM. However, we note that the Born-SOA difference is much less than the SOA-FEM difference; for attenuation at $2k_{0L}a = 1$, the former is -0.3% for aluminium and -14.9% for lithium.

In contrast to the low-frequency range, the middle-frequency range exhibits mostly smaller differences, and interestingly, the attenuation differences nearly overlap across the materials in the range of $2k_{0L}a = 5 - 12$ in Figure 4(a). At very high frequencies, the differences in both attenuation and phase velocity tend to grow with frequency and material anisotropy. It might be valuable to see how the differences progress with frequency in the future when computation power allows. Overall, the SOA model has a reasonable agreement with FEM in the middle- and high-frequency regions, and the single-scattering Born model agrees just as well with FEM even for strongly scattering polycrystals. This may suggest that multiple scattering is not strong even in these regions.

## (b) Dependence of attenuation and velocity on elastic scattering factors

It has been demonstrated by Figure 3 that attenuation and velocity variation increase with material anisotropy. A similar increase has also been found in Figure 4 for the relative difference between the theoretical and FE models, revealing a more prominent anisotropy dependence at relatively low and high frequencies. For this reason, two normalised frequencies, $2k_{0L}a = 1$ and $2k_{0L}a = 12$, are chosen in these two ranges to quantitatively evaluate this dependence. At these two frequencies, the normalised attenuation and phase velocity points are extracted from Figure 3 and plotted versus the elastic scattering factors in Figure 5. Since the two frequencies roughly fall into the Rayleigh and stochastic regimes, the elastic scattering factors $Q_{L \to T}$ and $Q_{L \to L}$ are employed to characterise the degrees of scattering respectively, see §3 for the selection of these factors. We note that the factor $\varepsilon^2 = \langle (k_L - k_{0L})^2 \rangle / k_{0L}^2$ defined in Refs. [5,39] is identical to the factor $Q_{L \to L}$ (thus refer to the $Q_{L \to L}$ column of Table 2 for the values of $\varepsilon^2$) that describes the degree of inhomogeneity for the longitudinal-to-longitudinal scattering in the stochastic regime; moreover, here we use the factor $Q_{L \to T}$ to describe the degree of inhomogeneity for the longitudinal-to-transverse scattering dominating in the Rayleigh regime [17,19].

At the low frequency $2k_{0L}a = 1$, the numerical FE points show a distinct quadratic relationship with the elastic scattering factor $Q_{L \to T}$ for both attenuation and phase velocity. Quadratic fits are generated for the data points and are plotted in the figure, and the fits are given by

$$2\alpha_L a = 44.52 Q_{L \to T}^2 + 0.28 Q_{L \to T}, \text{ or } \text{Im}k_L = 0.09/(2a) \times 4 Q_{L \to T}(0.78 + 4\pi^3 Q_{L \to T}) \quad (10)$$

$$V_L/V_{0L} - 1 = -61.69 Q_{L \to T}^2 - 2.77 Q_{L \to T}, \text{ or } \text{Re}k_L = k_{0L} + 2 Q_{L \to T}(1.39 + \pi^3 Q_{L \to T})k_{0L}, \quad (11)$$

with the goodness-of-fit of $R^2 = 0.997$ and $R^2 = 0.999$ respectively. By comparison, the theoretical SOA and Born predictions have a rather different dependence of linear order on the scattering factor $Q_{L \to T}$, which is represented by the linear fits in the figure. Also, these predictions are smaller in magnitude than the FEM points due to the aforementioned reason that the theoretical models only consider a subset of scattering events whereas the FEM considers all.

At the high frequency $2k_{0L}a = 12$, all results suggest a linear dependence on the scattering factor $Q_{L \to L}$, as is more evident from the attenuation results shown in Figure 5(b). Nonetheless, we emphasise that this assertion of linear dependence may not be appropriate. This is due to the possibility that not all evaluated materials are consistently in the stochastic regime at $2k_{0L}a = 12$ because weakly scattering materials are yet to enter the stochastic regime while the strongly scattering ones are already transiting into the geometric regime.

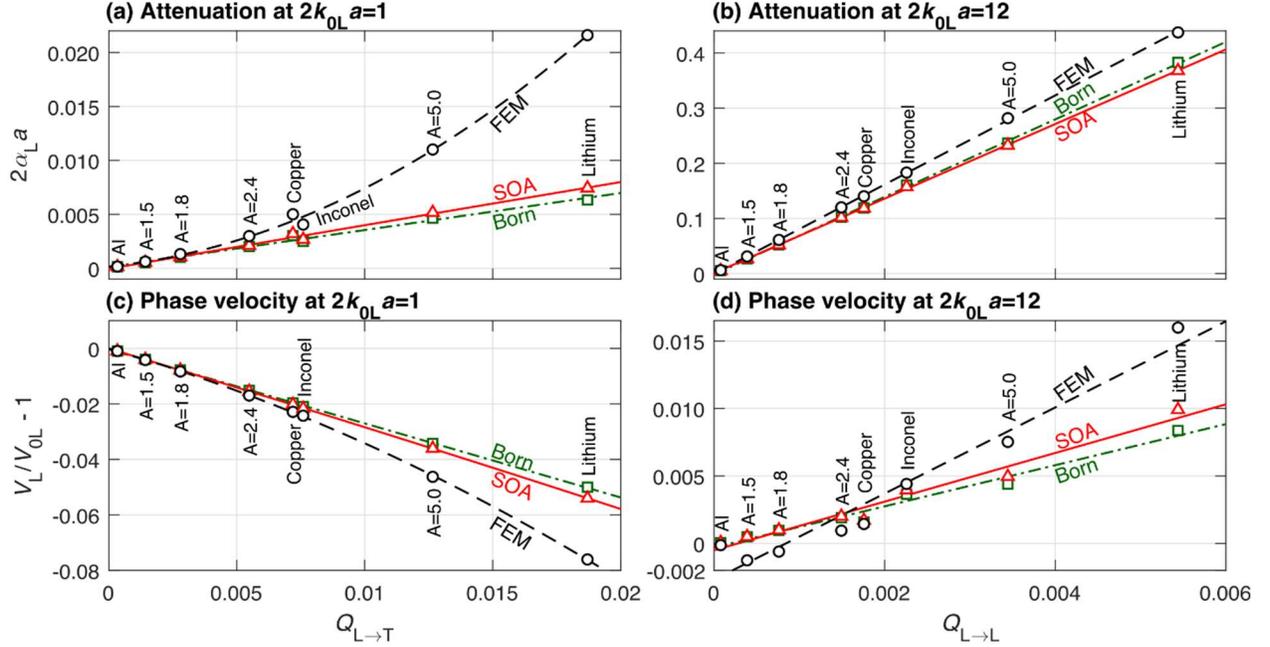

Figure 5. (a-b) Normalised attenuation and (c-d) phase velocity variation versus elastic scattering factors for longitudinal waves in cubic polycrystals with the PVT microstructure, comparing theoretical SOA and Born predictions with numerical FEM results. The FEM (circles), SOA (triangles) and Born (squares) points are taken from Figure 3 at the normalised frequencies of (a,c) $2k_{0L}a = 1$ and (b,d) $2k_{0L}a = 12$. Note that the y-axes of all four panels represent the attenuation and velocity variation induced by the total scattering (not by individual L → L or L → Tscattering components) while the x-axes use different elastic scattering factors depending on the normalised frequency. The dash lines in (a) and (c) are quadratic fits of the FEM points, while the rest lines are linear fits.

## (c) Quasi-static velocity limit

The quasi-static limit of longitudinal phase velocity is related to the effective elastic constant $C_{11}$ of the polycrystal medium by $V_L = \sqrt{C_{11}/\rho}$. Since this limit can be determined to a high degree of accuracy using 3D FEM [17], its FEM results can be utilised to evaluate the suitability of effective medium theories. For this reason, we calculate the quasi-static FEM velocities for the eight cubic materials with $A > 1$ and provide the results in Figure 6(a) as a function of the scattering factor $Q_{L \to T}$. Each point is statistically converged by taking the average of 30 realisations using the model N115200 (Table 1), and the respective error bar is not shown because it is smaller than the size of the markers.

The effective medium theories considered here include the Hashin-Shtrikman (HS) bounds [40–42] and the self-consistent (SC) theory [43,44]. The lower and upper HS bounds, denoted respectively as LHS and UHS, prescribe the limiting range for the quasi-static velocity, while the SC theory provides a unique estimation of quasi-static velocity satisfying the continuity of stress and strain throughout the polycrystal. The bounds in Figure 6(a) are calculated with the scripts by Brown [42] and the SC results with the scripts by Kube and De Jong [43]. Additionally, the Rayleigh asymptote of the SOA model, Eq. (8), is evaluated here and shown in the figure. The looser first-order bounds, namely the Reuss-Voigt bounds, are not shown, but note that all results in Figure 6(a) are normalised to the Voigt velocity, $V_{0L}$. The deviations of the theoretical predictions to the reference FE results are plotted in Figure 6(b). Quadratic fits are generated for all datasets shown in the figure.

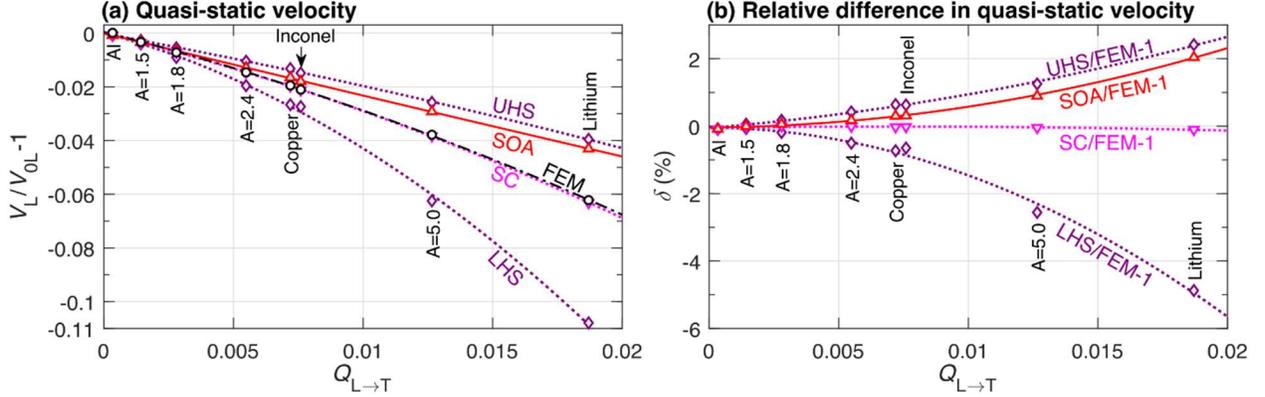

Figure 6. (a) Normalised quasi-static longitudinal velocity versus elastic scattering factor for cubic polycrystals with the PVT microstructure, comparing theoretical and numerical results; (b) relative difference between theoretical and numerical models versus elastic scattering factor. LHS and UHS represent respectively the lower and upper Hashin-Shtrikman bounds. SC denotes the self-consistent effective theory. SOA corresponds to the Rayleigh asymptote of the SOA model, Eq. (8). FEM represents the results obtained from quasi-static FE simulations and each point is the average of 30 simulations. The lines are the quadratic fits of the points. The SC and FEM points are nearly overlapped in (a).

Consistently for all materials, the FEM, SC and SOA points lie well between the LHS and UHS bounds, whose range becomes wider as the scattering factor increases. The SOA estimates perform well for weakly scattering materials, but the agreement deteriorates with the increase of grain anisotropy. The SC theory shows an excellent agreement with the FEM results even for lithium, while the SOA model is less satisfactory in this regard. All results show a quadratic relationship between the normalised quasi-static velocity $V_L/V_{0L} - 1$ and the elastic scattering factor $Q_{L \to T}$. This sole dependence on the elastic scattering factor is a new finding and is simpler than the previously observed dependence on both the Poisson's ratio and anisotropy index [16]. The differences between the theoretical and FEM results also exhibit a quadratic dependence on $Q_{L \to T}$.

## (d) Cubic materials with anisotropy indices $A < 1$

Here we present the results for the four cubic materials with anisotropy indices smaller than 1 ($A < 1$, Table 2). The results are provided in Figure 7, with panels (a) and (b) showing the normalised attenuation and phase velocity variation against the normalised frequency, (c) and (d) the respective results at $2k_{0L}a = 1$ versus the elastic scattering factor $Q_{L \to T}$, and (e) the quasi-static velocity against $Q_{L \to T}$. All shown FEM results are obtained by averaging the results of 15 realisations, while the calculations of the theoretical SOA curves are based on the generalised TPC function of the FE models, Supplementary Table 1. The FEM results are given only for the low-frequency Rayleigh range where our interest lies.

In contrast to the $A > 1$ case, a prominent finding is that the theoretical SOA predictions agree excellently with the FEM points. The normalised root-mean-square deviations (RMSD) of the SOA results from the FEM results are 0.94%, 4.57%, 5.48% and 6.88% for RbF, RbCl, RbBr and RbI for attenuation, while the respective numbers for phase velocity are 0.15%, 0.36%, 0.45% and 0.51%. We note that RbI has approximately the same $Q_{L \to T}$ as lithium and its universal anisotropy index [45] is of the same order of magnitude as that for lithium (2.64 for RbI and 8.70 for lithium). The two materials also show very similar levels of scattering, as can be observed from their FE wavefields, signals and wavefront fluctuations in Supplementary Figure 1. Despite these similarities, the SOA-FEM difference for RbI is an order of magnitude smaller than that for lithium. An in-depth study shows that the FEM results can be fitted to second-order polynomials of $Q_{L \to T}$ for both the $A < 1$ and $A > 1$ cases, but the second-order term for the $A < 1$ case is negligible while that for the $A > 1$ case is comparable to its linear-order term. By contrast, the SOA model predicts a linear dependence on $Q_{L \to T}$ for both cases. This contrasting result seems to be the reason why the SOA model performs well only for the $A < 1$ case while not for the $A > 1$ case.

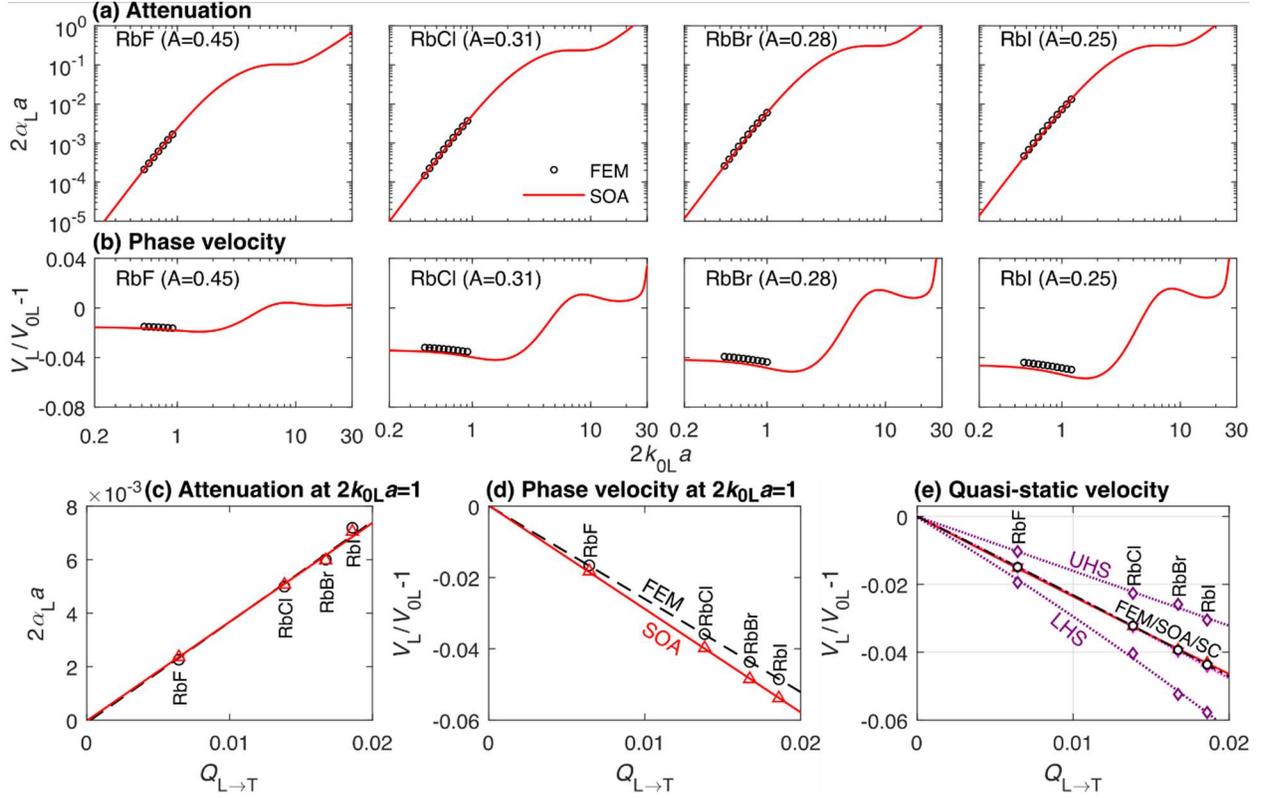

Figure 7. (a) Normalised attenuation and (b) phase velocity variation versus normalised frequency for plane longitudinal waves in four cubic polycrystals with the PVT microstructure and $A < 1$, comparing the SOA and FEM results. (c) and (d) show the normalised attenuation and velocity variation at $2k_{0L}a = 1$ against the elastic scattering factor $Q_{L \to T}$. (e) displays the velocity variation at the quasi-static limit.

Both $A > 1$ and $A < 1$ cases in Figure 6 and Figure 7(e) exhibit an exceptional agreement between the SC and the quasi-static FEM results. This fact further supports the generality of the above conclusions on the dependences of $Q_{L \to T}$ in the low-frequency range. We further confirmed this finding by analysing an extra set of 441 cubic materials (153 with $A < 1$ and 288 with $A > 1$, using elastic constants from Ref. [46]) by comparing our quasi-static results with those obtained from the SC estimate [43] and the Reuss bound; the details of this study will be reported elsewhere. In particular, we found from the simple analytical expression of the Reuss bound [47] that the quasi-static velocity bound has both linear and quadratic $Q_{L \to T}$ terms for either case; however, the quadratic term is very much smaller than the linear term for the $A < 1$ case, whereas the two terms are comparable for the $A > 1$ case. Therefore, there seem to be some general grounds for the difference between the $A < 1$ and $A > 1$ cases; however, it is not clear to us why, physically, the coefficient for the second-order $Q_{L \to T}$ term is so small for $A < 1$.

## 5. Semi-analytical model for strongly scattering polycrystals with $A > 1$

The above attenuation and velocity results show a reasonably good agreement between the theoretical SOA and numerical FE models at high frequencies beyond the Rayleigh regime. In the low-frequency Rayleigh regime, however, the results exhibit a contrasting feature for the $A < 1$ and $A > 1$ cases. This is because the FEM results have different dependences on $Q_{L \to T}$, being nearly linear, at $A < 1$, and quadratic, at $A > 1$, whereas the SOA model exhibits a linear order for both cases. This leads to an excellent agreement between the SOA and FE models for the $A < 1$ case but a less satisfactory agreement for the $A > 1$ case that deteriorates rapidly with the increase of $A$. For this reason and the fact that most structural materials have anisotropy indices greater than unity, our focus of this section will be on the $A > 1$ case only.

The SOA model is inherently approximate; among those approximations the most important are:

(1). The SOA model involves a major approximation by replacing a discrete polycrystal with a continuous random medium with fluctuating elastic tensor and statistical representation of the polycrystal by the TPC function [5,8–10]. This replacement is intuitively applicable to materials of weak anisotropy but may introduce non-negligible errors for strongly anisotropic materials.

(2). The SOA model uses the first-order smoothing approximation, as described by Weaver for polycrystals [8]. In the equivalent diagram perturbation series method, this relates to accounting for a subset of the scattering diagrams in the solution of the exact Dyson equation [9,10]. This approximation is also equivalent to the Keller approximation [7] as applied by Stanke and Kino [5] and in Ref. [6] to solids. The neglected scattering events may be negligible for weakly scattering materials but become increasingly important as material anisotropy gets stronger.

(3) The model assumes the validity of factorizing the two-point correlation function into the elastic and geometric terms [5,8]. There is some numerical support for the validity of this factorization [48].

(4) The high orders of the scattering diagrams depend on the multipoint correlation functions [9,10]. The effect of the additional statistics on scattering was not addressed in the literature.

The effect of those approximations on the obtained solution is not yet known even for the scalar case due to the lack of exact solutions. Therefore, numerical methods are the only alternative at this time to evaluate the quality of obtained solutions for polycrystals.

The fact that the FEM results depend quadratically on the elastic scattering factor $Q_{L \to T}$ indicates that an iterative approach may be applied to the theoretical SOA model (with a linear dependence on $Q_{L \to T}$) to add a higher-order term on the scattering factor. Following Rytov et al. [10] (pages 139-141), we may produce the iteration series for the SOA model by obtaining an initial effective wave number from the dispersion equation, then using it as the wave number for the reference medium to get from the dispersion equation the next iteration for the effective wave number. Further repeating this one obtains higher-iteration solutions. Evidently, one iteration is sufficient to introduce a quadratic term of the elastic scattering factors into the solution. However, we note that even an infinite iteration will only consider a summation of a subset of the infinite types of scattering diagrams [10], and the contribution of the unaccounted diagrams seems significant as revealed from comparison with the FEM results. Thus, it is not feasible for this approach to fully take into account all scattering events. Also, note that the continuous random medium approximation (1) is done even before the Dyson equation can be derived by this approach.

Alternatively, we propose a semi-analytical model by using the first iteration of the far-field Born approximation that results in a corrective second-order term on the scattering factor and then significantly increasing the coefficient of the corrective term for the model prediction to match the FEM results. Surprisingly, this empirical coefficient is nearly $\pi^3$, and as will be seen below this semi-analytical model works very well for various cubic materials with different microstructures and also for cubic materials with elongated grains [49]. We start the iteration by assuming that the effective wave number $k_L$ of the Born approximation, Eq. (5), becomes the wave number of the reference medium [10] and thus substituting this wave number into Eq. (5) to form a new effective wave number. Since we are dealing with the low-frequency range where the $L \rightarrow T$ scattering is dominant, we use only the $L \rightarrow T$ term in Eq. (5) for the iteration. We do the iteration separately for the attenuation and phase velocity based on Eqs. (6) and (7); this results in the expressions:

$$\alpha_L = \sum_i A_i \frac{4Q_{L\rightarrow L} k_{0L}(k_{0L}a_i)^3}{1+4(k_{0L}a_i)^2} + \sum_i A_i \frac{4Q_{L\rightarrow T}(1+4\pi^3 p_i^{Im} Q_{L\rightarrow T})k_{0L}(k_{0T}a_i)^3}{\left[1+(k_{0T}a_i)^2(\eta_{LT}^2-1)\right]^2 + 4(k_{0T}a_i)^2}, \tag{12}$$

$$\begin{aligned} \mathrm{Re}\,k_L = k_{0L} &+ \sum_i A_i \frac{2Q_{L\rightarrow L} k_{0L}(k_{0L}a_i)^2}{1+4(k_{0L}a_i)^2} + 2Q_{LL}^* k_{0L} \\ &+ \sum_i A_i \frac{2Q_{L\rightarrow T}(1+\pi^3 p_i^{Re} Q_{L\rightarrow T})k_{0L}(k_{0T}a_i)^2 \left[1+(k_{0T}a_i)^2(\eta_{LT}^2-1)\right]}{\left[1+(k_{0T}a_i)^2(\eta_{LT}^2-1)\right]^2 + 4(k_{0T}a_i)^2}. \\ &+ \sum_i 2A_i Q_{L\rightarrow T}\left(1+\pi^3 p_i^{Re} Q_{L\rightarrow T}\right)k_{0L} \end{aligned} \tag{13}$$

The terms $4p_i^{Im} Q_{L\rightarrow T}$ and $2p_i^{Re} Q_{L\rightarrow T}$ directly come from the iteration. The coefficients $\pi^3$ and $\pi^3/2$ are obtained by best matching Eqs. (12) and (13) with the FEM results at $2k_{0L}a = 1$, Eqs. (10) and (11). The iterative factors are given originally by $p_i^{Im} = p_i^{Re} = 1/\left\{\left[1+(k_{0T}a_i)^2(\eta_{LT}^2-1)\right]^2 + 4(k_{0T}a_i)^2\right\}$, but our parametric study indicates that they need to be modified as follows to improve the transition of the semi-analytical model into the stochastic regime

$$p_i^{Im} = \frac{1}{1+(\eta_{LT}^2+1)(k_{0T}a_i)^2}, \ \ p_i^{Re} = \frac{1}{1+\frac{1}{2}(\eta_{LT}^2+1)(k_{0T}a_i)^2}. \tag{14}$$

We note that the empirical coefficient $\pi^3/2$ for $\mathrm{Re}\,k_L$ (phase velocity) is half of that for attenuation, $\pi^3$, and there is also an extra $1/2$ constant in $p_i^{Re}$ as compared to $p_i^{Im}$. This systematic difference between the correction coefficients for attenuation and phase velocity is not yet understood. It is worth pointing out that removing the summations over $i$ in Eqs. (12) and (13) would deliver a semi-analytical model for polycrystals with the TPC given by a single exponential, $w(r) = e^{-r/a}$. Also, we emphasise that the empirical coefficients are nearly unchanged when the SBCs are accounted for because they only affect the TPC function and the effect is small, see §4(a) and Ref. [28].

**_Rayleigh asymptotes:_** At the Rayleigh limit, the attenuation and phase velocity asymptotes obtained from Eqs. (12) and (13) are given in the especially simple form

$$\alpha_{\mathrm{L}}^{\mathrm{R}} = \frac{1}{2\pi} k_{0\mathrm{L}}^4 V_{\mathrm{eff}}^{\mathrm{g}} \left[ Q_{\mathrm{L}\to\mathrm{L}} + \frac{V_{0\mathrm{L}}^3}{V_{0\mathrm{T}}^3} (1 + 4\pi^3 Q_{\mathrm{L}\to\mathrm{T}}) Q_{\mathrm{L}\to\mathrm{T}} \right], \quad V_{\mathrm{L}}^{\mathrm{R}} = \frac{V_{0\mathrm{L}}}{1 + 2Q_{\mathrm{LL}}^* + 2Q_{\mathrm{L}\to\mathrm{T}}(1 + \pi^3 Q_{\mathrm{L}\to\mathrm{T}})} \tag{15}$$

The stochastic attenuation and phase velocity asymptotes for the semi-analytical model are the same as those given by Eqs. (9) because the model has retained the same stochastic behaviours as the Born approximation.

Below, by comparing with the FEM results, we evaluate the applicability of the semi-analytical model, Eqs. (12) and (13), to different materials and microstructures and also assess its accuracy for the quasi-static velocity limit, Eq. (15).

**_Applicability of the semi-analytical model to cubic polycrystals with various anisotropy indices:_** First, we evaluate the applicability of the semi-analytical model to the eight cubic materials with the same PVT microstructure but greatly differing anisotropy indices. The semi-analytical model predictions are compared with the FEM and SOA results in Figure 3. The figure shows that the semi-analytical model mostly overlaps with the FEM at low frequencies for both attenuation and phase velocity across all materials, demonstrating its greatly improved accuracy in comparison to the SOA model for high grain anisotropies. Table 3 summarises the normalised RMSD of the models from the FEM (the FEM values as the reference). It reveals that the semi-analytical model mostly performs an order of magnitude better than the SOA model in the low-frequency range; the difference between the semi-analytical and FE models barely shows dependence on material anisotropy, especially for attenuation. Although not shown, we note that the Rayleigh asymptotes of the semi-analytical model also excellently represent the attenuation and velocity behaviours at the low-frequency limit. In the transition region, the semi-analytical model exhibits a slightly better agreement with the FEM results than the SOA model, and Table 3 suggests that the bettering of the agreement is more evident for materials of a stronger scattering. The semi-analytical model approaches the SOA model in the stochastic range for all shown cases.

**_Applicability of the semi-analytical model to different polycrystal microstructures:_** The excellent agreement between the semi-analytical model and the FEM discussed above are for the data sets for which the matching coefficients $\pi^3$ and $\pi^3/2$ were determined in the model, Eqs. (12) and (13). Nonetheless, it is also important to compare the model for unrelated microstructures. Here we evaluate the applicability of the semi-analytical model to different polycrystal microstructures with greatly contrasting TPC. Among the eight materials with $A > 1$, the above analysis illustrates that lithium most critically challenges the existing SOA and Born models. For this reason, the lithium polycrystal is utilised here for the evaluation and it is additionally simulated with the Laguerre and CVT microstructures, §2. The resulting FEM points and those for the PVT microstructure (already shown in Figure 3) are plotted in Figure 8, compared with the predictions of the SOA and semi-analytical models. In contrast to the SOA model, the semi-analytical model has a remarkably better agreement with the FEM in both the low-frequency and transition regions for all three microstructures. In the low-frequency region, in particular, the RMSD for the attenuation, given in Table 3, decreases from 60% for the SOA model to 5% for the semi-analytical model, while that of phase velocity reduces from 2% to 0.3%. The figure also shows that the semi-analytical model practically overlaps with the SOA model in the stochastic range but starts to deviate from the latter because the model is developed based on the Born approximation.

Essentially there is no difference in the model performance for those three microstructures, indicating the independence of the semi-analytical model on the TPC of polycrystals; however, obviously, the TPC should be accurately measured.

Table 3. Normalised RMSD of the SOA and semi-analytical (S-A) models with the FEM in the Rayleigh and transition regions.

| | | RMSD in the Rayleigh region (FEM as reference) | | | RMSD in the transition region (FEM as reference) | | |
|---|---|---|---|---|---|---|---|
| | | $2k_{01}a$ | SOA | S-A | $2k_{01}a$ | SOA | S-A |
| Attenuation | Aluminium | ≤1 | 8.02% | 4.24% | 1-10 | 7.14% | 5.01% |
| | A=1.5 | ≤1 | 14.26% | 1.27% | 1-10 | 9.43% | 3.96% |
| | A=1.8 | ≤1 | 19.81% | 3.02% | 1-10 | 12.66% | 4.15% |
| | A=2.4 | ≤1 | 28.98% | 9.35% | 1-10 | 17.83% | 5.48% |
| | Copper | ≤2 | 34.55% | 5.90% | 2-10 | 13.60% | 7.01% |
| | Inconel | ≤2 | 32.50% | 10.79% | 2-10 | 13.93% | 6.87% |
| | A=5.0 | ≤1 | 50.56% | 9.18% | 1-10 | 30.17% | 7.25% |
| | Lithium | ≤1 | 63.74% | 3.75% | 1-10 | 36.94% | 10.79% |
| | Lithium-Laguerre | ≤1 | 62.77% | 5.16% | 1-6 | 32.82% | 12.67% |
| | Lithium-CVT | ≤1 | 62.95% | 3.37% | 1-10 | 40.08% | 9.15% |
| Phase velocity | Aluminium | ≤1 | $2\times10^{-4}$% | $6\times10^{-4}$% | 1-10 | $9\times10^{-3}$% | $8\times10^{-3}$% |
| | A=1.5 | ≤1 | 0.01% | $3\times10^{-3}$% | 1-10 | 0.05% | 0.04% |
| | A=1.8 | ≤1 | 0.04% | $2\times10^{-3}$% | 1-10 | 0.07% | 0.04% |
| | A=2.4 | ≤1 | 0.14% | 0.01% | 1-10 | 0.14% | 0.07% |
| | Copper | ≤2 | 0.23% | 0.03% | 2-10 | 0.21% | 0.14% |
| | Inconel | ≤2 | 0.24% | 0.03% | 2-10 | 0.22% | 0.16% |
| | A=5.0 | ≤1 | 0.77% | 0.09% | 1-10 | 0.59% | 0.33% |
| | Lithium | ≤1 | 1.79% | 0.16% | 1-10 | 1.24% | 0.76% |
| | Lithium-Laguerre | ≤1 | 1.98% | 0.33% | 1-6 | 0.82% | 0.60% |
| | Lithium-CVT | ≤1 | 1.78% | 0.15% | 1-10 | 1.49% | 0.90% |

***Applicability of the semi-analytical model to quasi-static velocity limit:*** Finally, we appraise the applicability of the semi-analytical model to the quasi-static velocity limit, which may be of particular interest in developing effective medium theories. As shown in Figure 9(a), the semi-analytical model predictions can hardly be distinguished from the FEM points, and their relative difference shown in the inset is below 0.1% for all evaluated materials, achieving an order of magnitude improvement in accuracy in comparison to the SOA model. The normalised RMSD between the semi-analytical and FEM results over the eight materials is 0.04% whereas that between the SOA and FEM points is 0.81%. As shown in Figure 9(a), the SC estimate has the same excellent agreement with the FEM. Considering the additional results shown in Figure 6 and Figure 7 (e), it is reasonable to assume that the SC estimate would perform well for any cubic polycrystal. There are other reasons to believe that the SC method provides accurate estimates of homogenised elastic moduli of polycrystals: a) iterative convergence of high order bounds to the SC values [43] and b) FEM confirmation of the SC results [50]. For this reason, we use the SC estimate to calculate the quasi-static velocities for the same 288 materials with $A > 1$ ($A = 9.81$ is the largest) [46] as in §4(d), and then we use the SC results as the reference to further evaluate the applicability of the semi-analytical model. It is found that the semi-analytical results nearly overlap with the SC estimations, Figure 9(b), with the normalised RMSD over the 288 materials being 0.04% (this is the same as that between the semi-analytical and FEM results in Figure 9(a)). This further substantiates the validity of the semi-analytical model and also indicates the generality of the model constants, $\pi^3$ and $\pi^3/2$, for cubic materials with $A > 1$. We note that the SC theory in general needs to be iteratively solved [43], whereas the semi-analytical model has a simple explicit expression given by Eq. (15).

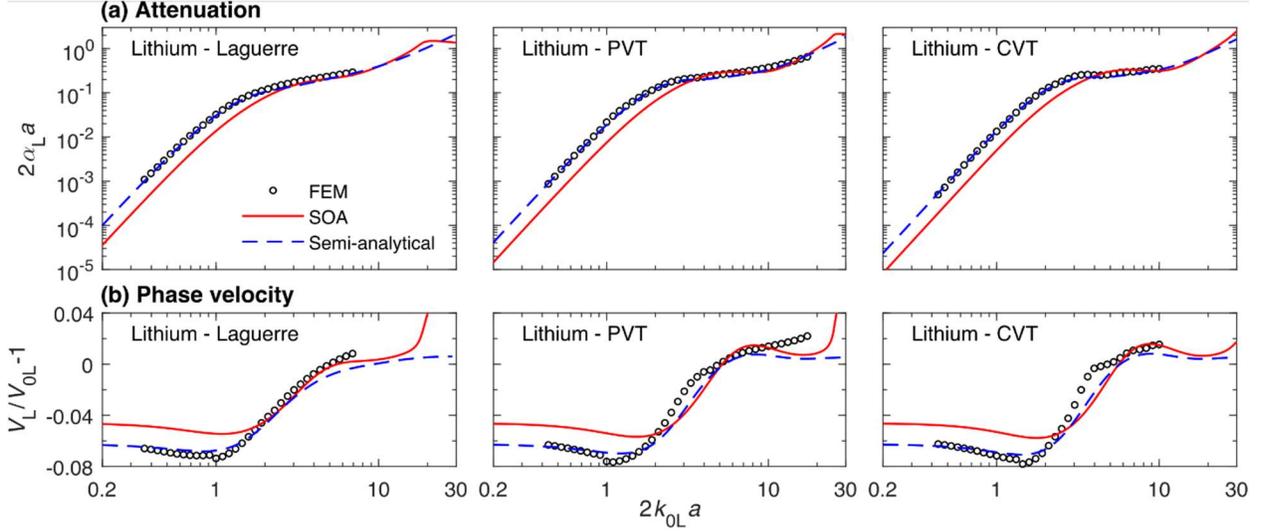

Figure 8. Normalised (a) attenuation and (b) phase velocity versus normalised frequency for plane longitudinal waves in polycrystals with statistically equiaxed grains of different uniformities, comparing theoretical SOA and semi-analytical predictions with numerical FEM results.

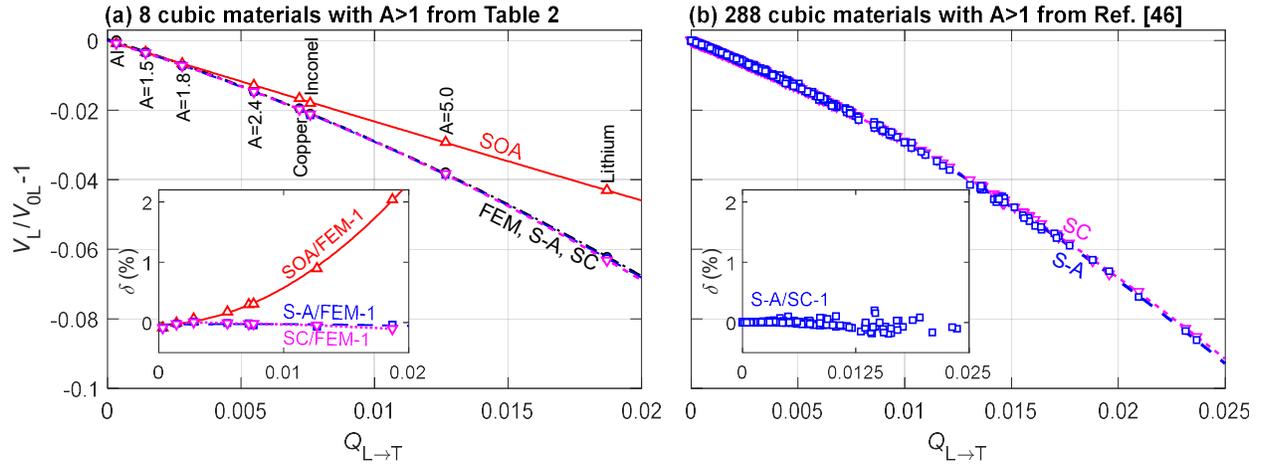

Figure 9. Normalised quasi-static longitudinal velocity versus elastic scattering factor for cubic materials with $A > 1$. (a) compares the FEM, SOA and SC results (taken from Figure 6) with the semi-analytical results (S-A, calculated using Eq. (15)) for the 8 cubic materials with $A > 1$ from Table 2. The inset in (a) shows the respective relative differences of the model predictions to the FEM results, and those for S-A and SC are within ±0.1%. (b) compares the S-A and SC results of 288 cubic materials with $A > 1$ from Ref. [46]. The inset in (b) displays the respective relative difference, which is within ±0.2%; the normalised RMSD over all materials is 0.04%. Lines are linear/quadratic fits.

## 6. Summary and conclusions

This work uses 3D FE and theoretical models to study the scattering-induced attenuation and phase velocity variation of plane longitudinal waves in untextured cubic polycrystals with statistically equiaxed grains. The study is predominantly performed for materials with anisotropy indices greater than unity ($A > 1$). The results of such materials exhibit a good agreement between the SOA and FE models in the transition and stochastic regimes, even for very strongly scattering lithium ($A = 9.14$). This agreement also holds for the single-scattering Born approximation, thus indicating the possibility that the effect of multiple scattering on the coherent wave is weak in these regions. In the low-frequency Rayleigh regime, the theoretical models agree reasonably well with the FEM for common structural materials with $A < 3.2$: the largest difference in attenuation between the SOA and FEM is -10%, -35% and -37% for aluminium, Inconel and copper (the figures are slightly larger for the Born-FEM difference). However, the relative difference can reach the level of -70% for more strongly scattering materials like lithium. The emergence of such unsatisfactory agreement in the Rayleigh regime for cubic materials with $A > 1$ is somewhat unexpected.

A study on materials with $A < 1$ is also conducted in the Rayleigh regime. It shows an excellent agreement between the SOA and FE models, with an attenuation difference of smaller than 7% for RbI that has nearly the same elastic scattering factor $Q_{L \to T}$ as lithium. Further analysis reveals that this excellent agreement is due to the nearly linear dependence of the FEM results on $Q_{L \to T}$ (proportional to the elastic covariance), which is the same as for the SOA results. By contrast, the FEM results for the $A > 1$ case are described by a quadratic polynomial on $Q_{L \to T}$ with a significant dependence on the $Q_{L \to T}^2$ term, while the SOA model results are still linear to $Q_{L \to T}$ (irrespective of $A$ being larger or smaller than one). This contrasting dependence leads to the unsatisfactory SOA-FEM agreement for the $A > 1$ case. In addition to the FEM evidence, the linear and quadratic dependences of the $A < 1$ and $A > 1$ cases on $Q_{L \to T}$ are supported by the quasi-static velocity limit, particularly by the self-consistent estimate and Reuss bound. The SOA model is inherently approximate, and its disagreement with the FEM may be attributed to the replacement of the polycrystal by a continuous random medium and by approximations in the solution of the Dyson equation, mainly by limiting the order of perturbations, thus not accounting for all scattering events.

To consider strongly scattering materials with $A > 1$, we have proposed a semi-analytical model by iterating the far-field Born approximation and optimising the coefficient of the second-order term on the scattering factor $Q_{L \to T}$ to achieve the best fit of the model to the FE results. We have demonstrated that the semi-analytical model works remarkably well for all materials considered in this work and for different polycrystal microstructures with largely differing TPC. The largest difference in attenuation between the semi-analytical and FE models is within a reasonable ±15% range for all evaluated materials and microstructures. The semi-analytical model also delivers a very accurate prediction for the quasi-static velocity limit obtained by the FEM. In addition to the FE evidence supporting the semi-analytical model to predict the quasi-static velocity limit, an excellent agreement is observed between the semi-analytical model and the self-consistent estimate for 288 materials with $A > 1$ (with a normalised RMSD of 0.04%, which is within the accuracy of the FE method). This finding substantiates the generality of the empirical semi-analytical model coefficients $\pi^3$ and $\pi^3/2$ for cubic materials with $A > 1$.

The applicability of the proposed model is demonstrated for materials of cubic crystal symmetry, but we expect that the model may be applicable to polycrystals of lower symmetries and general inhomogeneous materials after the adjustment of the iterative coefficients. We hope that the simple form of the proposed semi-analytical model and its exceptional performance against the FE simulations will stimulate more rigorous theoretical developments.